\documentclass{article}
\usepackage{spconf,amsmath,graphicx}
\usepackage{amsmath}
\usepackage{amsthm}
\usepackage{framed,multirow}
\usepackage{graphicx}
\usepackage{placeins}
\usepackage{color}
\usepackage{booktabs}
\usepackage{enumitem}
\setlist{nosep, leftmargin=14pt}

\usepackage{mwe} % to get dummy images

\title{Hierarchical Agent-based Reinforcement Learning Framework for Automated Quality Assessment of Fetal Ultrasound Video}
\name{Sijing Liu \qquad Qilong Ying \qquad Shuangchi He \qquad Xin Yang \qquad Dong Ni \qquad Ruobing Huang} 
\address{Medical Ultrasound Image Computing (MUSIC) Lab, Health Science Center, Shenzhen University, China}

\begin{document}
%\ninept
%
\maketitle
\begin{abstract}
%The abstract should contain about 100 to 150words.
Ultrasound is the primary modality to examine fetal growth during pregnancy, while the image quality could be affected by various factors. Quality assessment is essential for controlling the quality of ultrasound images to guarantee both the perceptual and diagnostic values. Existing automated approaches often require heavy structural annotations and the predictions may not necessarily be consistent with the assessment results by human experts. Furthermore, the overall quality of a scan and the correlation between the quality of frames should not be overlooked. In this work, we propose a reinforcement learning framework powered by two hierarchical agents that collaboratively learn to perform both frame-level and video-level quality assessments. It is equipped with a specially-designed reward mechanism that considers temporal dependency among frame quality and only requires sparse binary annotations to train. Experimental results on a challenging fetal brain dataset verify that the proposed framework could perform dual-level quality assessment and its predictions correlate well with the subjective assessment results.

\end{abstract}
\section{Introduction}
\label{sec:intro}
Real-time sonography is the most widely used imaging technique for prenatal evaluation \cite{salomon2011practice}. It is economical, painless, and non-ionizing, while the image quality can be user-dependent and thereby may affect subsequent diagnosis and management \cite{frederiksen2012advances}. Multiple factors affect the acquired images, including the skills and experience of the operator, fetal position, gestational age, etc. Clinical guidelines were published to help control the ultrasound(US) image quality and promote the detection of abnormalities \cite{salomon2006feasibility}. They defined a series of standardized planes to evaluate key anatomical structures and acquire reproducible biometric measurements. This process requires specialized knowledge and a comprehensive understanding of the rapidly changing fetal anatomy and the corresponding sonographic patterns. 

Efforts have been made to automate this process. Existing methods usually transform the task of US quality control into a structure detection problem on 2D planes \cite{wu2017fuiqa,dong2019generic}. This reveals the location and the size of certain structures through the `eye' of a detection model, while is not equivalent to whether this structure is of its optimal visibility (i.e., high quality) to the naked eye. As a result, their results may not be consistent with the subjective assessment rated by human experts. Fluctuated or inaccurate detection predictions are also expected which can lead to incorrect quality assessment (QA) results. Furthermore, the training of these models necessitates detailed annotations (e.g., bounding boxes of structures in every plane), which is labor-intensive and time-consuming. Innovative methods that are label-efficient and more in line with experts' QA ratings are therefore needed. 

Another limitation of current popular approaches is that they often assess the quality of a single US frame separately (e.g. classifying it into whether it is a standardized plane or not), ignoring the fact that US examination is a dynamic process. In other words, they built upon the assumption that the quality of frames or planes is independent of each other and fail to incorporate the rich temporal features that lie within a video. Furthermore, it is desired to have tools that could highlight videos of overall inferior quality to assist the less experienced operators to improve scan quality. 

To address these, we design a novel framework by evaluating the frame-level and video-level quality simultaneously during a fetal ultrasound examination. The main contributions are:\\
1. A reinforcement learning (RL) based framework that jointly considers the quality of each frame and the whole video. The proposed tool not only provides standard planes but also calls attention to scans of unsatisfying quality that may require a re-scan.\\
2. A novel reward mechanism that considers the temporal correlation between the quality of consecutive frames. It also allows direct exploitation of subjective quality assessment scores and does not require detailed structural annotations. \\ 
3. A hierarchical agent design consists of a subordinate agent that performs frame-level quality assessment (FQA) and a superordinate agent that performs video quality assessment (VQA). This allows flexible feature extraction in different spatial or temporal dimensions while enabling collaborative learning that benefits both tasks.

\section{Related works}
\label{sec:format}
\textbf{US quality assessment.}
The quality control/assessment of fetal US images is indispensable to obtaining reproducible measurements and accurate diagnosis. To achieve this, Wu et al. used a classification CNN model to generate probability maps of key regions and classify the ROIs~\cite{2-1}. 
Dong et al. proposed a two-staged framework to classify the cardiac four-chamber planes of cardiac US and score the quality of each plane based on whether a certain structure can be detected~\cite{2-2}. In \cite{2-4}, Lin et al. used a multi-task model to detect six anatomical structures and also scored the quality of each image based on the number of detectable structures. Similarly, Zhang et al. combined a region proposal network and a classification model to classify whether a US image is qualified \cite{2-5}. \\
\textbf{Standard plane detection.} 
Identifying the standard planes is one of the most important steps during a prenatal US screening. The quality of the obtained planes is essential to the evaluation of fetal growth and thus is closely related to the QA task for fetal US. To address this, Baumgartner et al. designed a 6-layer CNN model which allowed robust scan plane detection~\cite{1-1}. Lei et al.proposed a multi-task learning model that jointly optimized key structure recognition and plane classification tasks for fetal head ultrasound \cite{1-8}. Chen et al. combined LSTM and CNN to automatically detect several standard planes \cite{1-3}. Pu et al. proposed an automatic standard plane recognition model using a two-branch model with optical flow~\cite{1-7}. Liu et al. utilized the Bi-LSTM model for sequential modeling under a reinforcement learning framework to generate a summarization of US videos~\cite{1-4}. 

 \begin{figure*}[!htb]
	\centering
	\includegraphics[width=0.68\linewidth]{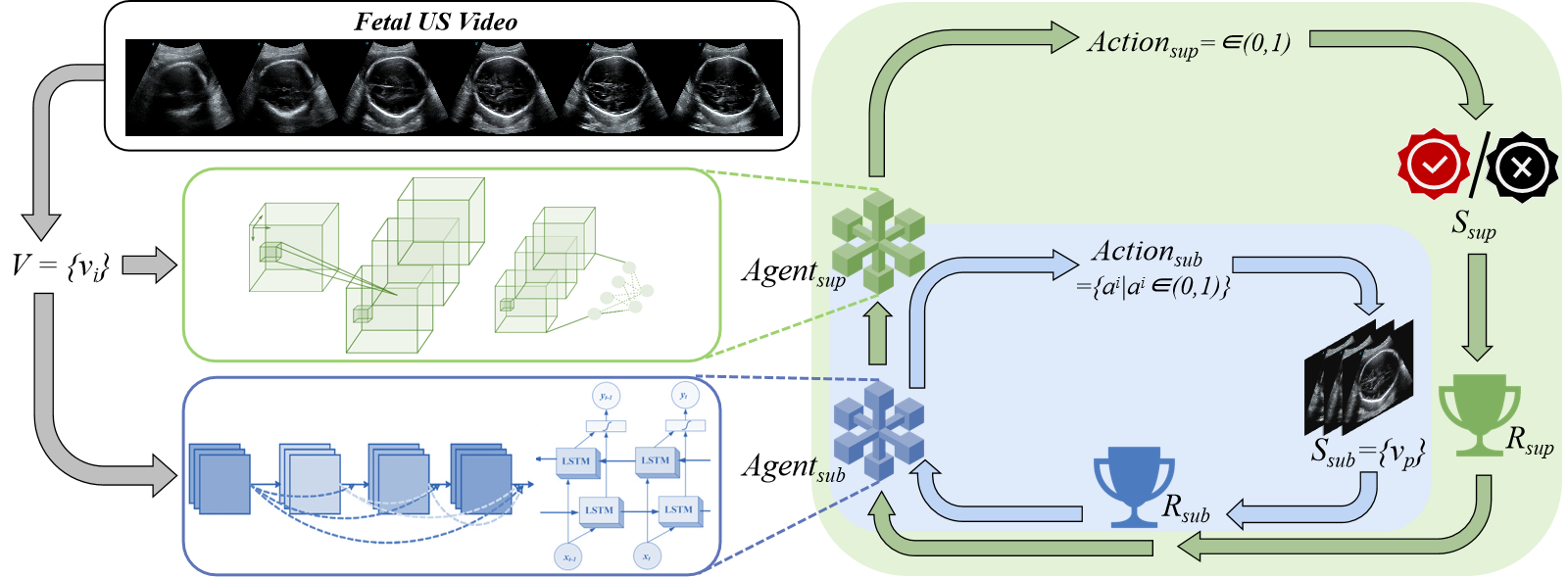}
	\caption{The proposed hierarchical agent-based RL framework. It consists of an inner loop (blue) where $Agent_{sub}$ performs frame-level quality assessment, and an outer loop (green) where $Agent_{sub}$ assesses the overall video quality based on the input video and the features learned by $Agent_{sub}$.}
	\label{fig:framework}
\end{figure*}

\section{Method}
\label{sec:method}
 The quality of US images reveals both whether it is clinically valuable (e.g., relevant anatomical structures are visible) and whether it is visually satisfying to human viewers. This highly abstract task aims to find the mapping from the pixel space to the perceptual assessment scores rated by human experts. Moreover, the quality of a single image and that of a whole US scan are correlated, yet intrinsically different, as the latter should take the overall spatial-temporal information into account. 
 To better coordinate the learning between the two, we integrate them through the RL-based model with hierarchical agents. The whole pipeline is shown in Fig.\ref{fig:framework}.
 
This framework follows the common setting of RL with slight alterations: two agents in their respective state $S_{sub}$, $S_{sup}$, interact with the environments $E$ by making suitable action $action_{sub}$, $action_{sup}$ to maximize the expectation of the total reward $R_{total}$. Formally, a raw US video with unfixed length can be denoted as $V=\{v^{i}\}_{1}^{N}$, with the size of $H \times W \times C \times N$, while N represents the number of frames. We also define the following elements:

\textbf{Agents:}
The proposed framework consists of a subordinate agent $Agent_{sub}$ that processes a US video to perform frame selection and a superordinate agent $Agent_{sup}$ that takes both the US video and the features of $Agent_{sub}$ to assess the overall quality of the video. Both of them sample sequentially from a Bernoulli distribution to generate suitable actions. In specific, $Agent_{sub}$ is composed of a 2D CNN model (e.g., DenseNet) to extract spatial features of each frame $v^i$, followed by a one-layer Bi-LSTM model (with a hidden size of 256) to further combine them and yield the high-level features $f_f=LSTM(Dense(V))$. Finally, it outputs the probability for each frame that should be selected as `qualified' using a sigmoid layer. Contrarily, $Agent_{sup}$ utilizes the classical C3D model to extract spatial-temporal features of the whole video $f_v$. $f_v$ and then fused with $f_f$ to perform the video quality rating. $Agent_{sub}$ is first pre-trained to warm up the learning. The two agents are then trained together end-to-end to allow interactive learning. This hierarchical design enables flexible feature learning with different focuses, but also guides the video quality rating task with the frame-level quality information. \\
\textbf{Action:}
The two agents have different action space.
$Action_{sub}$ is defined as the set of frame-wise QA operation: $Action_{sub} = \{a^i|a^i \in \{0,1\} \}_1^{N}$, where `1' indicates the current frame is `qualified' and should be selected in the standard plane set, while `0’ indicates otherwise. $Action_{sup}$, instead, is a video-level operation: $Action_{sup}  \in \{0,1\} $, where `1’ indicates satisfactory video quality, `0’ indicates otherwise.\\
\textbf{State:}
State $S_{sub} = \{v_{p}\}_1^{N_{p}}$ corresponds to the standard plane set that is selected by $Agent_{sub}$. $N_p$ denotes the total number of selected frames.
State $S_{sup}  = \hat{Q}_v $ denotes the quality of the whole video predicted by $Agent_{sup}$.\\
\textbf{Reward:}
One key limitation of existing QA approaches fails to consider the temporal dependency of the quality of adjacent frames. In other words, the quality of the consecutive frames of a `qualified' frame is also expected to be high, and vice versa. Meanwhile, analyzing fetal US videos faces unique challenges as multiple `clusters' of qualified planes may exist within a single video as different structures may be identified. Note that these clusters may overlap with each other or be far apart given different fetal positions, probe manipulation trajectory/speed, and spatial relationships among different organs. To handle these without requiring explicit structural annotations, we leverage help from the signal processing theory to smooth the binary QA results rated by human experts. Given the ground truth frame-level quality score for the whole video: $Q_f = \{q_f^i|q_f^i \in \{0,1\} \}_1^{N}$, it can be expressed as a combination of several rectangular pulse functions, each of which is defined by $Rect(\tau_t,\Delta\tau_t)$, where $\tau_t$ defines the activation time point while $\Delta\tau_t$ defines its activation interval. Formally, $Q_f =\sum_t^T{Rect(\tau_t,\Delta\tau_t)}$, where $T$ is the total number of such pulses. The reward for $Agent_{sub}$ can be calculated as:
\begin{equation}
\resizebox{.45 \textwidth}{!} 
{$R_{sub} = env(Tr(id|\tau_1,\Delta\tau_1),Tr(id|\tau_2,\Delta\tau_2),..,Tr(id|\tau_T,\Delta\tau_T))$}
\end{equation}
\begin{equation}
\resizebox{.48 \textwidth}{!} 
{$
Tr(id|\tau_t,\Delta\tau_t) = \left\{
\begin{array}{lcl}
\displaystyle \frac{A_{max}}{d}(id-\tau_t+d)  & & { \tau_t -d \leq id \leq \tau_i}\\
\displaystyle A_{max} & & { \tau_t \leq id \leq \tau_t+ \Delta\tau_t}\\
\displaystyle \frac{A_{max}}{d}(\tau_i+\Delta\tau_t+d-id)  & & { \tau_t+\Delta\tau_t \leq id 
      \leq \tau_t+\Delta\tau_t+d} \\
\displaystyle -1 & & else
\end{array}\right.
$}
\end{equation}
, where $env()$ denotes the envelope function, $Tr()$ represents a trapezoidal wave function, $d$ and $A_{amx}$ are hyperparameters that control the slope and the amplitude of the trapezoid. $Tr()$ could be considered as a temporally-smoothed $Rect()$, that proportionally credits the frames that are adjacent to high-quality ones based on their distance. We also introduce an extra negative reward of $-1$ to penalize the agent if unqualified frames were selected to avoid trivial solutions. The $env()$ function handles the clusters overlapping scenario and could process videos with arbitrary numbers and distribution of qualified frames. 

$Agent_{sup}$ is trained using a cubic reward function to penalize inaccurate video quality prediction $\hat{Q}_v$, denoted as:
\begin{equation}
R_{sup} = {-||\hat{Q}_v-Q_v||}^3
\end{equation}
 The total reward can be represented as $ R_{total} = R_{sub}+\beta*R_{sup}$, where $\beta$ is a hyper-parameter that controls the weighting of the two tasks.
\section{Experiments}
\label{sec:exp}
\textbf{Dataset.}
We evaluate the proposed approach using a US fetal brain dataset collected at Shenzhen Maternity\&Child Healthcare hospital. The study was approved by the local Institutional Review Board. It contains 878 US videos of different fetuses, with a gestational age range from 29 to 40 weeks. Note that this dataset is especially challenging to analyze due to skull calcification and rapid brain development during the third trimester. All videos were then resized to have a height of 256 and a width of 256, and split or padded to have a fixed length of 128 for easier analysis. To simplify the annotation process, one clinical expert ($>$8 years of experience) was invited to rate the quality of each frame into two classes (i.e., `poor'(0) or `satisfactory'(1)), based on whether this frame could be used in routine examination. Similarly, the expert also rated the overall quality of each video into either `qualified' or `unqualified' based on whether this video is of acceptable quality for standard examination or a re-scan is required after full evaluation of the whole video.\\ 
\textbf{Experiments.}
The model performance is evaluated using standard accuracy, precision, specificity, sensitivity, and F1-score in both frame- and video-level. We compared the proposed model with both the state-of-the-art approaches in frame-level QA (i.e., T-RNN~\cite{1-3}, UVS~\cite{1-4}, FUSPR~\cite{1-7}, 3DST-UNet~\cite{1-9}). 3DST-UNet and UVS also utilized an RL-based framework. Note that some frame-based QA methods such as~\cite{2-1} are not feasible here as no structural annotation was available for this study. As there exists no established baseline for VQA of fetal US, we also implement 2 strong video DL models as the competing methods (i.e., C3D~\cite{tran2015learning}, VST~\cite{liu2022video}).
Another factor worth investigating is whether the proposed dual-agent design is beneficial for both tasks. To investigate this, we carry out an ablation study by removing the $Agent_{sup}$. All quantitative results are shown in Table 1.\\
\textbf{Implementation.}
All experiments were implemented in PyTorch with a GeForce RTX 3090 GPU. For the RL-based algorithm, the number of episodes is fixed to 5. An SGD optimizer with momentum is used to train the models. The initial learning rate is set to $1e^{-5}$ and subsequently reduced by a factor of 0.5 for every 30 epochs. 
\section{Results and discussion}
%\subsection{Comparison experiments}
%\label{ssec:subhead}
\FloatBarrier
\begin{table*}[!hbt]
    \centering
    \caption{Experimental results of the quality assessment accuracy in both frame- and video-level.}
    \label{tab:quantitative}   
    \resizebox{0.90\textwidth}{!}{%
    \begin{tabular}{@{}|c|ccccccc|ccccccc|@{}}
    \hline
    \multirow{2}{*}{Method}  &  \multicolumn{7}{c|}{Frame Quality Assessment} & \multicolumn{7}{c|}{Video Quality Assessment} \\\cline{2-15}                                             
    & ACC(\%)        & SEN(\%)        & SPE(\%)        & PRE(\%)     & F1(\%)  & AUC(\%)  & $p<0.05$  & ACC(\%)        & SEN(\%)       & SPE(\%)        & PRE(\%)   & F1(\%)  & AUC(\%) & $p<0.05$\\\hline
UVS        & $87.79$          & $62.33$          & $93.41$             & $67.63$          & $64.87$  & $77.87$  & $Yes$
&   --     &    --      &  --    &    --    & --   & -- & --\\
3DST-UNet  & $83.23$          & $45.61$        & $91.53$              & $54.34$          & $49.60$       & $68.57$  & $Yes$
&   --     &  --      &  --     &    --    & --    & --  & -- \\ 
FUSPR      & $85.93$          & $69.45$     & $89.57$      & $59.52$    & $64.10$       & $79.50$    & $Yes$
&   --   & --     &    --      &  --      &    --    & --    & -- \\ 
T-RNN   & $87.62$    & $41.19$    & \boldmath{$97.87$}    & \boldmath{$81.03$}   & $54.62$    & $69.52$  & $Yes$
&   --   & --     &    --    &  --     &    --      & --   &    --\\ 
\hline \midrule                                     
C3D    &  --      &   --    &    --      &  --     &    --    & --   &    --  & $86.67$  & $89.20$  & $83.06$   & $88.20$   & $88.70$   & $86.13$ & $Yes$\\
VST    &  --   &   --    &    --  &  --     &    --    & --   &  -- 
& $58.67$  & $100$   & $0$  & $58.67$   
& $73.95$   & $50.0$  & $Yes$\\
\hline \midrule                                     
Ours w/o $R_{sup}$        
 & $88.69$          & $68.97$      & $93.05$       & $68.66$          & $68.81$    & $81.01$   & $Yes$ 
&   --    &  --      &    --     &  --     &    --      & --  & --  \\
Ours  & \boldmath{$88.80$}    & \boldmath{$71.37$}     & $92.61$     & $68.07$   & \boldmath{$69.68$}    & \boldmath{$81.99$} & --  
  & \boldmath{$90.67$}    & \boldmath{$93.75$  }    & \boldmath{$86.29$}    & \boldmath{$90.66$ } 
  & \boldmath{$92.18$ }     & \boldmath{$90.02$ } & -- \\
\hline
    \end{tabular}
}
    \end{table*}
Table \ref{tab:quantitative} shows that the proposed framework performed superiorly in frame-level QA than the competing methods in ACC, SEN, F1-score, and AUC, indicating a balanced performance. Statistical tests further proved that the difference in AUC between the proposed and all its variants is significant (delong test, $p<0.05$). This suggests that our method produces QA results that are consistent with those rated by human experts. Note that the T-RNN scored high SPE and PRE scores, while its SEN and F1-score are substantially lower than ours. This may be caused by severe overfitting to the `unqualified' class. On the contrary, our model obtained promising performance in all metrics, showing its ability to handle the class imbalance between qualified planes and the unqualified ones. The UVS model also scored a relatively good ACC and AUC, which are still lower than those of the proposed model. This may be the result of the proposed reward mechanism that considers the temporal correlation of the quality among adjacent frames. Furthermore, the UVS model does not penalize false-positive predictions, while the proposed $R_{sub}$ explicitly considers this through the additional negative reward. Note that the 3DST-UNet also shares a similar RL setting, while it yielded slightly lower performance. This may be explained by that the frame-level QA decision-making relies more on 2D spatial information, thus the 3D-based model is not suitable for this task.

Row 5-6 of Table.\ref{tab:quantitative} display the VQA performance of different approaches. To the best of our knowledge, this work is the first to perform QA for both the frames and the whole video jointly for fetal US data. Experimental results prove its versatility, which exceeded popular baselines (C3D and VST) in video analysis. It is also interesting to see that the VST scored inferiorly to C3D model despite its sophisticated design and the incorporation of a multi-layer self-attention mechanism and suitable hyper-parameter tuning. We conjecture that the VST's performance might be improved given a larger dataset due to its significantly larger model size.

The penultimate row of Table.\ref{tab:quantitative} reports the results of the ablation study where $Agent_{sup}$  (i.e. outer loop in Fig.\ref{fig:framework}) is removed from the proposed framework. It can be seen that the incorporation of video-level quality information can help to boost the performance in frame-level QA as well, especially in SEN. Note that the performance of the C3D model also indicates how the proposed model would perform without $Agent_{sub}$, as $Agent_{sup}$ also utilizes the same C3D architecture. It can be seen that the $Agent_{sup}$ benefited substantially from the subordinate frame-based QA task, and obtained a 4\% increase in ACC, 4.55\% increase in SEN, and 3.89\% increase in AUC. This may stem from that the $Agent_{sub}$ is trained with frame-level supervision that may be more informative than the single binary quality label used by $Agent_{sup}$. As a result, the latter gained more from the former during the training, while co-learning still proved to be advantageous as both of the two tasks reported higher performance in all metrics.

% Subheadings should appear in lower case (initial word capitalized) in
% boldface.  They should start at the left margin on a separate line.
 \begin{figure}[!htbp]
	\centering
	\includegraphics[width=0.98\linewidth]{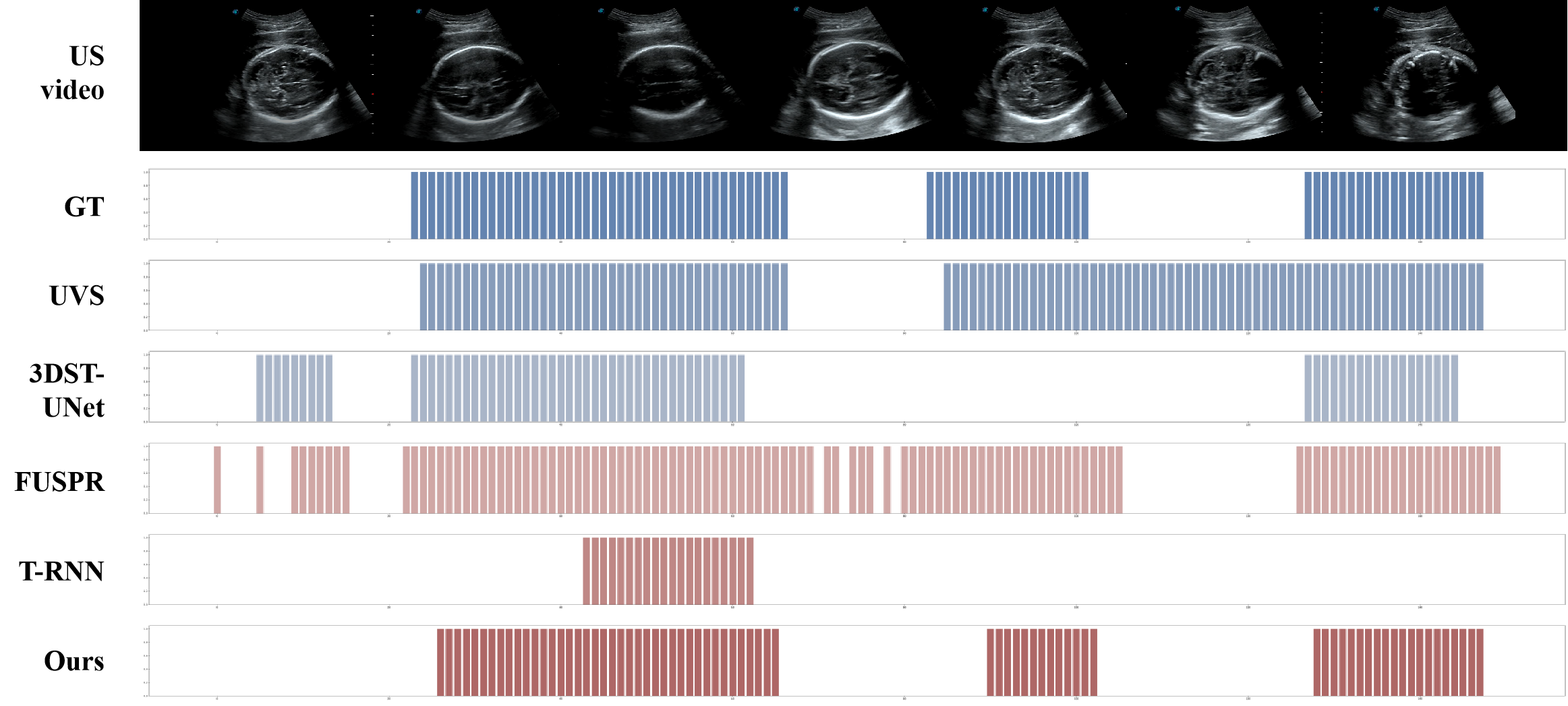}
	\caption{Visual results of the comparison experiment. }
	\label{fig:results}
\end{figure}

Figure \ref{fig:results} provides a visual example of the comparison experiments. A test data is randomly selected (row 1). As a result, the quality of frames exhibits a complex pattern (GT QA score shown in dark blue, row 2). Row 3-7 reveal how competing methods performed in frame-level QA. Overall, the UVS model's predictions are close to the GT, while being over-confident in part of the scan. FUSPR model rated most frames as `qualified', leading to more false-positive results (low SPE). Fast probe movement might have caused larger difference among frames and may have distracted the model. On the contrary, 3DST-UNet and T-RNN chose fewer frames as standard planes (i.e., qualified frames). 
This suggests that these two either have problems in capturing long-range temporal dependency or might have overfitted the training data. Contrarily, the proposed framework excelled in this challenging case and its predictions align well with the GT (row 7).

\section{Conclusion}
\label{sec:print}
In this paper, we proposed an RL framework that is equipped with hierarchical agents for FQA and VQA for fetal US scans. It can automatically yield standard planes and also highlight videos with overall unsatisfying quality. The framework has been evaluated on a challenging dataset of fetal brain, while the overall methodology is general and could be applied to analyze other datasets. Additionally, it is worth noting that this pioneering study is the first to address automated VQA for fetal US while it only considers data collected from the same center. Future studies may also explore whether it is suitable for data collected in multiple sites and the impact of US vendors on VQA.

\section{Acknowledgments}
\label{sec:acknowledgments}
This work was supported by the National Natural Science Foundation of China (No. 62101342, and No. 62171290); Guangdong Basic and Applied Basic Research Foundation (No.2023A1515012960); Shenzhen-Hong Kong Joint Research Program (No. SGDX20201103095613036)).
% Print your properly formatted text on high-quality, 8.5 x 11-inch white printer
% paper. A4 paper is also acceptable, but please leave the extra 0.5 inch (12 mm)
% empty at the BOTTOM of the page and follow the top and left margins as
% specified.  If the last page of your paper is only partially filled, arrange
% the columns so that they are evenly balanced if possible, rather than having
% one long column.

%\section{Acknowledgments}
%\label{sec:acknowledgments}

% References should be produced using the bibtex program from suitable
% BiBTeX files (here: strings, refs, manuals). The IEEEbib.bst bibliography
% style file from IEEE produces unsorted bibliography list.
% ------------------------------------------------------------------------- 
\bibliographystyle{IEEEbib}
\bibliography{strings,refs}

\end{document}